\documentclass{ws-procs9x6}

\newcommand{\vecc}[1]{\mbox{\boldmath $#1$}}

\begin{document}
\hfill{RUB-TPII-06/2010}

\title{TRANSVERSE-MOMENTUM PARTON DENSITIES:\\
GAUGE LINKS, DIVERGENCES AND SOFT FACTOR\footnote{Talk presented at
Workshop on ``Exclusive Reactions at High Momentum Transfer'', 18-21 May 2010,
TJNAF (Newport News, VA, USA)}}

\author{I. O. CHEREDNIKOV$^\ddag$\footnote{Also at: ITPM, Moscow State University, Russia}}

\address{INFN Cosenza, Universit$\grave{a}$
             della Calabria \\  I-87036 Rende (CS), Italy\\
             and
             \\
             Bogoliubov Laboratory of Theoretical Physics \\
             JINR
             RU-141980 Dubna, Russia\\
$^\ddag$E-mail: igor.cherednikov@jinr.ru}

\author{N. G. STEFANIS$^\S$}

\address{Institut f\"{u}r Theoretische Physik II, \\
             Ruhr-Universit\"{a}t Bochum,
             D-44780 Bochum, Germany\\
$^\S$E-mail: stefanis@tp2.ruhr-uni-bochum.de}

\begin{abstract}
We discuss the state-of-the-art of the theory of transverse-momentum
dependent parton densities (TMD)s, paying special attention to their
renormalization properties, the structure of the gauge links in the
operator definition, and the role of the soft factor in the
factorization formula within the TMD approach to the semi-inclusive
processes.
We argue that the use of the lightcone axial gauge offers certain
advantages for a consistent definition of TMDs as compared to the
off-the-light-cone gauges, or covariant gauges with off-the-lightcone
gauge links.

\end{abstract}


\bodymatter

\section{Introduction}

The distribution functions of partons (in what follows we consider
only quark distributions), depending on the longitudinal components
$x$, as well as on the transverse components $\vecc k_\perp$, of their
momenta (hence TMD)s, accumulate useful information about the intrinsic
motion of the hadron's constituents and enter as a nonperturbative
input in the QCD approach to the semi-inclusive hadronic processes
(see, e.g., Refs. [\refcite{Sop77, Col78, Col03, BR05}]).
The QCD factorization formula for a semi-inclusive structure function
is expected to have the following symbolic form
[\refcite{CS81, CS82, JMY04}]
\begin{equation}
  F  (x_B, z_h, \vecc P_{h\perp}, Q^2)
  =
  \sum_i \ e_i^2 \cdot
 {H }  \otimes {{\cal F}_D}
  \otimes {{\cal F}_F} \otimes {S} \ ,
  \label{eq:factor_symb}
\end{equation}
where $z_h$ and $\vecc P_{h\perp}$ are the longitudinal and transverse
fractions of the momentum of the produced hadron, respectively.
This expression contains the hard (perturbatively calculable) part $H$,
the (nonperturbative) distribution and fragmentation functions
${{\cal F}_D}$ and
${{\cal F}_F}$, and the soft part $S$.
The latter is absent in the collinear (fully inclusive) picture, and
will be discussed below.
However, several problems arise in attempting to formulate the TMD
approach in terms of the quantum field operators and their matrix
elements:
$(i)$ Extra (rapidity) divergences appear already at the one-loop
level, which invalidate the standard renormalization procedure
[\refcite{Col03, Col08, CS07, CS08}].
$(ii)$ A much more complicated (compared to the collinear case)
structure of the gauge links leads to the non-universality of
distribution or fragmentation functions
(see, e.g., Refs. [\refcite{BoM07, CRS07, Bacch08}]).
$(iii)$ Several counter-examples have been given showing that the
straightforward factorization formula (\ref{eq:factor_symb}) may fail,
at least in some specific situations [\refcite{CQ07, RM10}].
$(iv)$ The role and explicit expression of the soft factor $S$ can be
different in different schemes.
In what follows, we basically concentrate on the first and the last
problem.

\section{Divergences and renormalization properties of TMDs}
\label{aba:sec1}

The operator definition of the quark TMD (without the soft term) reads
[\refcite{CS81, CS82, CS07, CS08, CH00, JY02, BJY02, BMP03, CM04, Hau07}]
\begin{eqnarray}
  && \tilde {\cal F}_{i/h} \left(x, \mbox{\boldmath$k_\perp$}\right)
= \frac{1}{2}
  \int \frac{d\xi^- d^2\mbox{\boldmath$\xi_\perp$}}{2\pi (2\pi)^2}
  {\rm e}^{-ik^{+}\xi^{-} +i \mbox{\footnotesize\boldmath$k_\perp$}
  \mbox{\footnotesize\boldmath$\xi_\perp$}} \nonumber
\\
&&
  \times \left\langle
              h |\bar \psi_i (\xi^-, \mbox{\boldmath$\xi_\perp$})
              [\xi^-, \mbox{\boldmath$\xi_\perp$};
   \infty^-, \mbox{\boldmath$\xi_\perp$}]_{[n]}^\dagger
   [\infty^-, \mbox{\boldmath$\xi_\perp$};
   \infty^-, \mbox{\boldmath$\infty_\perp$}]_{[\vecc l]}^\dagger
\right. \nonumber \\
&& \left. \times
      \gamma^+[\infty^-, \mbox{\boldmath$\infty_\perp$};
   \infty^-, \mbox{\boldmath$0_\perp$}]_{[\vecc l]}
   [\infty^-, \mbox{\boldmath$0_\perp$}; 0^-,\mbox{\boldmath$0_\perp$}]_{[n]}
   \psi_i (0^-,\mbox{\boldmath$0_\perp$}) | h
   \right\rangle \, .
\label{eq:tmd_naive}
\end{eqnarray}
This expression may be given a physical meaning, because it is formally
gauge invariant.
The gauge invariance is ensured by means of the path-ordered gauge
links
\begin{eqnarray}
\begin{split}
  & [\infty^-, \mbox{\boldmath$\xi_\perp$}; \xi^-, \mbox{\boldmath$\xi_\perp$}]_{[n]}
\equiv {}
 {\cal P} \exp \left[
                     i g \int_0^\infty d\tau \ n_{\mu}^- \
                      A_{a}^{\mu}t^{a} (\xi + n^- \tau)
               \right] \, , \\
               &
 [\infty^-, \mbox{\boldmath$\infty_\perp$};
 \infty^-, \mbox{\boldmath$\xi_\perp$}]_{[\vecc l]}
\equiv {}
 {\cal P} \exp \left[
                     i g \int_0^\infty d\tau \ \mbox{\boldmath$l$}
                     \cdot \mbox{\boldmath$A$}_{a} t^{a}
                     (\mbox{\boldmath$\xi_\perp$}
                     + \mbox{\boldmath$l$}\tau)
               \right] \, ,
\end{split}
\end{eqnarray}
where we distinguish between longitudinal
(lightlike, $n^2=0$) $[...]_{[n]}$
and transverse $[...]_{[\vecc l]}$
links.
(A generalized definition, which includes into the Wilson
lines the spin-dependent Pauli term
$F^{\mu\nu}[\gamma_\mu, \gamma_\nu]$, was recently worked out
in Ref.\ [\refcite{CKS10}]).

Of course, the above expressions have to be quantized, using, for
instance, functional-derivative techniques.
This means that the gluon potential in the gauge link has to be
Wick contracted with corresponding terms in the interaction Lagrangian,
accompanying the Heisenberg fermion (quark) field operators.

At the tree-level, the ``distribution of a quark in a quark''
(here we consider only ultraviolet (UV) and rapidity divergences,
which are independent of the particular hadronic state) is normalized
as
\begin{eqnarray}
  && \tilde {\cal F}_{q/q}^{(0)} (x, \vecc k_\perp) =
  \frac{1}{2}
   \int \frac{d\xi^- d^2\xi_\perp}{2\pi (2\pi)^2}
   {\rm e}^{- i k^+ \xi^- + i k_\perp \cdot \xi_\perp}
  \nonumber \\
  &&
   \times { \langle p | }\bar \psi (\xi^-, \xi_\perp)
  \gamma^+
   \psi (0^-,0_\perp) {| p \rangle } =
  \delta(1 - x ) \delta^{(2)} (\vecc k_\perp) \ ,
  \label{eq:tree_tmd}
\end{eqnarray}
and, formally, the integration over $\vecc k_\perp$ yields the usual
collinear (integrated) PDF
\begin{equation}
\int\! d k_\perp^{(2)} \tilde {\cal F}_{q/q}^{(0)} (x, \vecc k_\perp)
=
F_{q/q}^{(0)} (x) = \delta(1 - x ) \ .
\end{equation}
However, already in the calculation of the one-gluon contributions, one
encounters---besides the normal UV divergences---certain pathological
singularities.
Namely, one has at the one-loop level the following singular terms:
\begin{enumerate}

  \item Standard UV poles ${\sim \frac{1}{\varepsilon}}$ in the
  dimensional regularization:
  they can be removed by the usual $R-$operation
  and are controlled by renormalization-group evolution
  equations (analoguous to the DGLAP equation in the integrated case).

  \item Pure {\it rapidity divergences}: they give rise to logarithmic
  and double-logarithmic terms of the form
  ${\sim \ln \eta \ , \ \ln^2 \eta }$.
  These terms, although they depend on the additional
  rapidity parameter $\eta$ [\refcite{Sop77, CS81, CS82, Col08}],
  do not affect the UV renormalization properties and can be safely
  resummed, e.g., by means of the Collins-Soper equation.

  \item Pathological {\it overlapping divergences}: they contain the UV
  and rapidity poles simultaneously
  $
  {\sim \frac{1}{\varepsilon} \ \ln \eta \ }
  $
  and are considered to be {\it highly problematic}.
  The reason is that they prevent the removal of {\it all}
  UV-singularities by the standard $R-$procedure.
  Therefore, a special {generalized} renormalization procedure is needed
  in order to take care of those terms and enable the construction of
  well-defined renormalizable TMDs.

\end{enumerate}
It is interesting to note that working in the lightcone gauge with the
Mandelstam-Leibbrandt prescription [\refcite{Man83, Lei84}], one
doesn't get any overlapping divergences, at least in the leading loop
order [\refcite{CS09}].
The renormalization properties of operators and matrix elements
containing Wilson lines and loops with or without obstructions have
been extensively studied in various situations---see, e.g., Refs.\
[\refcite{Pol79, CD80, Ste83, KR87, BB89}].
The specifics of the TMD consist in the fact that, though the fermion
fields are separated by a spacelike distance, the gauge links lay on
pure lightlike rays, or on the $2D-$transverse plane at lightcone
infinity.

The analysis of the one-loop anomalous dimension of the TMD, given by
Eq.\ (\ref{eq:tmd_naive}), shows that the contribution of the
overlapping singularity is nothing else, but the {\it cusp anomalous
dimension} [\refcite{KR87}].
Therefore, in order to renormalize expression (\ref{eq:tmd_naive}),
one can apply, apart from the standard $R-$operation, an additional
renormalization factor, which depends on the cusp angle and can be
written as a vacuum matrix element of the Wilson lines evaluated
along a special contour with an obstruction (cusp); viz.,
\begin{equation}
   Z_{\chi}^{-1}
   =
   \left\langle 0 \left| {\cal P} \exp \Bigg[ ig \int_{\chi}d\zeta^\mu
   \ t^a A^a_\mu (\zeta) \Bigg] \right| 0 \right\rangle \ .
\end{equation}
The UV singularity of this factor cancels the cusp anomalous dimension
from the overlapping divergence, thus rendering the re-defined TMD
(\ref{eq:tmd_naive}) renormalizable [\refcite{CS07, CS08}].
Therefore, the generalized renormalization procedure for the TMD can be
formulated as
\begin{equation}
   {\tilde {\cal F}_{\rm ren} (x, \vecc k_\perp, \chi, ...)
   =
    Z_{\rm R} \cdot Z_{\chi} \cdot \tilde {\cal F} (x, \vecc k_\perp, \chi, ...) } \ ,
\end{equation}
where $Z_{\rm R}$ is the usual renormalization constant, while $Z_{\chi}$
can be included in the definition of the TMD itself.
In that case, it is treated as a ``soft factor'':
\begin{equation}
  Z_{\chi} \equiv
  {\rm [Soft \ Factor]} \ ,
\end{equation}
which is defined as
\begin{eqnarray}
&& {\rm [Soft \ Factor]}
= \nonumber \\
  && { \langle 0
  |  } \ {\cal P}
  {\rm e}^{ig \int_{\mathcal{C}_{\chi}}\! d\zeta^\mu
           \ t^a A^a_\mu (\zeta)
      } \cdot
  {\cal P}^{-1}
  {\rm e}^{- ig \int_{\mathcal{C'}_{\chi}}\! d\zeta^\mu
           \ t^a A^a_\mu (\xi + \zeta)
      }
  {| 0
  \rangle } \ ,
\end{eqnarray}
where the contours ${\cal C}$ and ${\cal C'}$ are explicitly given in
Ref. [\refcite{CS08}].
Therefore, the generalized definition of the TMD reads
[\refcite{CS07, CS08, CS09}]
\begin{eqnarray}
  && {\cal F}_{i/h} \left(x, \mbox{\boldmath$k_\perp$}\right)
= \frac{1}{2}
  \int \frac{d\xi^- d^2\mbox{\boldmath$\xi_\perp$}}{2\pi (2\pi)^2}
  {\rm e}^{-ik^{+}\xi^{-} +i \mbox{\footnotesize\boldmath$k_\perp$}
\cdot \mbox{\footnotesize\boldmath$\xi_\perp$}}
\nonumber \\
&&
  \times \left\langle
              h \Big|\bar \psi_i (\xi^-, \mbox{\boldmath$\xi_\perp$})
              [\xi^-, \mbox{\boldmath$\xi_\perp$};
   \infty^-, \mbox{\boldmath$\xi_\perp$}]_{[n]}^\dagger
   [\infty^-, \mbox{\boldmath$\xi_\perp$};
   \infty^-, \mbox{\boldmath$\infty_\perp$}]_{[\vecc l]}^\dagger
\right. \nonumber \\
&& \left. \times
      \gamma^+[\infty^-, \mbox{\boldmath$\infty_\perp$};
   \infty^-, \mbox{\boldmath$0_\perp$}]_{[\vecc l]}
   [\infty^-, \mbox{\boldmath$0_\perp$}; 0^-,\mbox{\boldmath$0_\perp$}]_{[n]}
   \psi_i (0^-,\mbox{\boldmath$0_\perp$}) \Big| h
   \right\rangle
\nonumber \\
   && \times  \ {\rm [Soft \ Factor]}
   \, .
\label{eq:tmd_soft_f}
\end{eqnarray}

This function is free (at least, at the one-loop order) of pathological
divergences and is a well-defined renormalizable quantity.

\section{Factorization and role of the soft factor}

The soft factor, introduced above, naturally enters in the
factorization formula (\ref{eq:factor_symb}):
${\rm [Soft \ Factor]} = S$.
However, its interpretation is twofold.

On the one hand,  it formally looks similar to the ``intrinsic''
Coulomb phase found by Jakob and Stefanis [\refcite{JS91}] in QED
for Mandelstam charged fields involving a gauge contour which is
a timelike straight line.
The name ``intrinsic'' derives from the fact that this phase is
different from zero even in the absence of external charge
distributions.
Its origin was ascribed in [\refcite{JS91}] to the long-range
interaction of the charged particle with its oppositely charged
counterpart that was removed ``behind the moon'' after their
primordial separation.
Note that the existence of a balancing charge ``behind the moon''
was postulated before by several authors---see [\refcite{JS91}]
for related references---in an attempt to restore the Lorentz
covariance of the charged sector of QED.
This phase is acquired during the parallel transport of the charged
field along a timelike straight line from infinity to the point of
interaction with the photon field and is absent in the local approach,
i.e., for local charged fields joined by a connector.
It is different from zero only for a Mandelstam field with its own
gauge contour attached to it and keeps track of its full history
since its primordial creation.
Keep in mind that the connector is introduced ad hoc in order to
restore gauge invariance and is not part of the QCD Lagrangian.
In contrast, when one associates a distinct contour with each quark
field, one, actually, implies that these Mandelstam field variables
should also enter the QCD Lagrangian (see [\refcite{JS91}] for more
details).
However, a consistent formulation of such a theory for QCD is still
lacking and not without complications of its own.

The analogy to our case is the following.
First, formally adopting a direct contour for the gauge-invariant
formulation of the TMD in the light-cone gauge, the connector gauge
link does not contribute any anomalous dimension---except at the
endpoints; this anomalous dimension being, however, irrelevant for
the issue at stake.
Hence, there is no intrinsic Coulomb phase in that case.
Second, splitting the contour and associating each branch to a
quark field, transforms it into a Mandelstam field and, as a result,
adding together all gluon radiative corrections at the one-loop order,
a $\eta$-dependent term survives that gives rise to an additional
anomalous dimension.
We have shown that this extra anomalous dimension can be viewed as
originating from a contour with a discontinuity in the four-velocity
$\dot x(\sigma)$ at light-cone infinity---a cusp obstruction.

Classically, it is irrelevant how the two distinct contours
are joined, i.e., smoothly or by a sharp bend.
But switching on gluon quantum corrections, the renormalization
effect on the junction point reveals that the contours are not
smoothly connected, but go instead through a cusp [\refcite{CS08}].
Here, we have a second analogy to the QED case discussed above.
Similarly to the ``particle behind the moon'', this cusp-like junction
point is ``hidden'' and manifests itself only through the
path-dependent phase after renormalization.

In that case, the soft factor looks like an intrinsic property of the
gauge-invariant operators containing the fermion fields and must be
taken into account in order to construct consistently the gauge-invariant
renormalizable two-particle matrix elements.

On the other hand, the soft factor appears as the result of the
separation the ``soft'' contributions from the one-loop graphs
[\refcite{JMY04}]. In this situation it is needed in order to
avoid the double-counting in the factorization formula
(\ref{eq:factor_symb}). In general, these two soft factors might
be different. In particular, it has been shown in Ref.
[\refcite{CS09_er}] that the anomalous dimensions of the TMD
within different subtraction schemes of the soft factors are
different. The relationship between these frameworks will be
studied elsewhere.

\section{Conclusions}
\label{sec:concl}
The above mentioned results have been obtained by using the lightcone
axial gauge with a proper regularization of the gluon propagator
[\refcite{CS08, CS09}].
The extra rapidity divergences can be treated by different methods,
e.g., one may shift the gauge links off the lightcone, or use,
alternatively, the off-the-lightcone axial gauge
[\refcite{CS81, CS82, JMY04}].
In these cases, the additional rapidity variable parameterizes the
deviation of the gauge links from the light rays in terms of
the axial-gauge fixing vector.
An advantage of the ``pure lightcone'' frameworks is the more
straightforward physical interpretation of the factorization
and the role of the collinear Wilson lines in the definition of
TMDs, as well as the direct relationship between the (unintegrated)
TMDs and the (integrated) collinear PDFs: one can get the collinear
PDF, satisfying the DGLAP evolution equation, by simple
$\vecc k_\perp$-integration.
In contrast, the ``off-the-lightcone'' frameworks don't allow us to
perform such a procedure, so that more sophisticated methods must be
invented.
The complete proof of the QCD factorization, within the TMD approach
(in particular within the ``pure lightcone'' scheme), as well as the
clarification of the role played by the soft factors in different
approaches, are still lacking.


\end{document}